\documentclass[aps,prd,showpacs,notitlepage,nofootinbib,superscriptaddress,floatfix,showkeys,twocolumn]{revtex4-1}
\usepackage{float}
\usepackage{siunitx}
\usepackage{graphicx}
\usepackage{gensymb}
\usepackage{amsmath, cases}
\usepackage[usenames]{color}
\usepackage{amssymb}
\usepackage{hyperref}
\newcommand{\be}{\begin{equation}}
\newcommand{\ee}{\end{equation}}
\newcommand{\bea}{\begin{aligned}}
\newcommand{\eea}{\end{aligned}}
\newcommand{\pr}{\partial}

\newcommand{\bse}{\begin{subequations}}
\newcommand{\ese}{\end{subequations}}

\newcommand{\del}{\nabla}

\newcommand{\xb}{\bf x}
\newcommand{\yb}{\bf y}
\newcommand{\kb}{\bf k}
 
\newcommand{\pd}[2]{\frac{\partial #1}{\partial #2}} 
\renewcommand{\v}[1]{\ensuremath{\mathbf{#1}}} 
\renewcommand{\d}[2]{\frac{d #1}{d #2}} 
\newcommand{\bmm}{\begin{multline}}
\newcommand{\emm}{\end{multline}}

\newcommand{\mi}{\mathrm{i}}
\begin{document}
\title{Sonoluminescence: Photon production in time dependent analog system}
\author{Rajesh Karmakar}
\email{rajesh018@iitg.ac.in}
\affiliation{Department of Physics, 
Indian Institute of Technology Guwahati, Assam 781039, India}
\author{Debaprasad Maity}
\email{debu@iitg.ac.in}
\affiliation{Department of Physics, 
Indian Institute of Technology Guwahati, Assam 781039, India}


\begin{abstract}
Sonoluminescence is a well known laboratory phenomenon where an oscillating gas bubble in the appropriate environment periodically emits a flash of light in the visible frequency range. In this work, we study the system in the framework of analog gravity. We model the oscillating bubble in terms of analog geometry and propose a non-minimal coupling prescription of the electromagnetic field with the geometry. The geometry behaves as an analogous oscillating time dependent background in which repeated flux of photons are produced in a wide frequency range through parametric resonance from quantum vacuum. Due to our numerical limitation, we could reach the frequency up to $\sim 10^5 ~\mbox{m}^{-1}$. However, we numerically fit the spectrum in a polynomial form including the observed frequency range around $\sim 10^7 ~\mbox{m}^{-1}$. Our current analysis seems to suggest that parametric resonance in analog background may play a fundamental role in explaining such phenomena in the quantum field theory framework.

\end{abstract}

\maketitle

\newpage
\newpage
\section{Introduction}
In this paper, we propose a model of non-perturbative photon production in an analogue system. We build up the formalism considering a very special laboratory system consisting of an oscillating gas bubble immersed in water. 
This system has been extensively studied experimentally \cite{Gaitan, Hiller:1992qz, Gompf, Weninger, Camara} and observed to emit repeated flashes of photon flux, popularly known as Sonoluminescence in the literature. Despite the significant effort put in over the years the origin of the photon flux is still not understood completely. 

Emission from hot ionized gas in the framework of the classical hydrodynamical model has been one of the extensively discussed mechanisms to explain the phenomena. Due to the rapidly collapsing bubble, the gas inside is assumed to be partially ionized caused by the adiabatic heating. During the process, the produced accelerating free electrons may emit thermal photons through the Bremsstrahlung process. This was first proposed by Bradley P. Barber et.al.\cite{Barber}. Later on, such a mechanism has been further investigated in detail \cite{Lohse1999, Brenner:2002zz, Hammerprl, Hammer309, Hammer303, CY2002}. In this hydrodynamical formalism, there are  Speculations \cite{Bishwajyoti} as to whether a nonlinearly collapsing bubble would be able to account for the focussing of the shockwave into such a micrometre length scale, that is required for the Bremsstrahlung to happen.

A purely quantum mechanical approach was first adopted by J. Schwinger. Using the idea of dynamical Casimir Effect\cite{casimir1948}, which applies between two dynamical boundaries in the quantum field theory framework, J. Schwinger first constructed the formalism of photon production \cite{schwinger1992a,schwinger1992b,schwinger1993a,schwinger1993b,schwinger1993c, schwinger1993d, schwinger1994a} under the instantaneous collapse approximation. Following the same direction S. Liberati et al. \cite{Liberati:1998wg, Visser:1998bqu, Liberati:1999jq, Liberati:1999uw} further extended the formalism and calculated photon spectrum in terms of the Bogoliubov coefficients modelling the oscillating bubble in terms of time dependent refractive index for an infinite homogenous dielectric medium. In a somewhat different approach, Eberlin \cite{Eberlein:1995ex, Eberlein:1995ev} used the idea of the Unruh effect \cite{Unruh:1976db} and 
modelled the surface of the collapsing bubble as an accelerating mirror. The photon flux was generated by solving the Schrodinger equation applying time dependent perturbation theory. However, to match the experimental photon spectrum the velocity of the bubble surface turned out to be superluminal, and perturbative approximation may not be applicable as pointed out in \cite{Brevik:1998zs, Unnikrishnan:1996zz}. The subsequent studies, therefore, \cite{Milton:1996wm, Milton:1997ky, Unnikrishnan:1996zz, Lambrecht:1996rb} have raised questions on the appropriateness of the approach of the formalisms based on the dynamical Casimir effect, which may not be able to properly account for sonoluminescence effect.

Important to note that in all the stated quantum mechanical formulations (also look at \cite{Chodos:1998jr}) except the model based on the Bremsstrahlung process, typical photon energy flux turned out to be divergent in wave number. Hence adhoc cut-off has been introduced to account for the phenomena. However, Milton later argued (\cite{Milton:1996wm, Milton:1997ky}) that such divergent contributions in the quantum theory framework must vanish upon renormalization.  

In this paper, we will again stress the quantum mechanical production discussed before. 
Theoretical models earlier discussed can be viewed in the framework of quantum field theory where fields, such as photons, are excited from the quantum vacuum due to time dependent background. We argue that in the perturbative framework, the vacuum production usually comes with $k^2$ divergence if the conformal property of the electromagnetic (EM) field is not properly taken into account. In the conformal frame, however, we show that it is the non-perturbative parametric resonance which may be an appropriate mechanism that can explain the sonoluminescence.
We utilize the well known analog spacetime formalism \cite{Unruh:1980cg} taking into account the actual temporal evolution of the bubble surface by solving the well known Rayleigh-Plesset (RP)\cite{Rayleigh, Plesset, Barber} equation. We couple the Ricci scalar curvature, $\mathcal{R}$, of the analog geometry with an EM field preserving Lorentz symmetry following the idea that has been widely used in cosmology. This scalar curvature acts as a conformal breaking factor leading to the production of photons, which is otherwise absent in conformal space time \cite{Birrell:1982ix}. 

Though we do not claim to explain sonoluminescence in the observed frequency range \cite{Gaitan, Hiller:1992qz, Gompf, Weninger}, present theoretical results, however, seem to suggest that our framework for quantum mechanical particle production in time dependent analog spacetime may well be an interesting avenue to explore. Due to our present numerical limitation, we could not reach the actual frequency range of observation. However, we certainly obtained the photon spectrum in the lower frequency with reasonable magnitude which could be extrapolated to match the observation.

The rest of the paper has been assembled as follows: We first discuss the characteristics of sonoluminescence phenomena in Sec.\ref{sono_exp} and study the dynamics of the air bubble in water. Next, we construct the analog background geometry representing the fluctuation in the fluid in Sec.\ref{analog_metric}. With the nonminimal coupling, built upon the Ricci scalar curvature of the analog metric, we have demonstrated the canonical quantization procedure in Sec.\ref{quantize}. Subsequently, we have derived the expression for spectral number density in terms of the EM fields and presented our results characterizing the growth of the number density, by which we have interpreted the photon production. In Sec.\ref{energyflux} we have evaluated the energy flux of the emitted photons from the number density and compared it with the experimental results. In Sec.\ref{highfreq_flux}, our analytical estimation in high frequency regime justifies the need to follow the formalism for nonperturbative production. Finally, we have concluded with a future outlook. 

\section{Sonoluminescence: experimental results}\label{sono_exp}
\begin{figure}[t]
\centering
\includegraphics[scale=0.6]{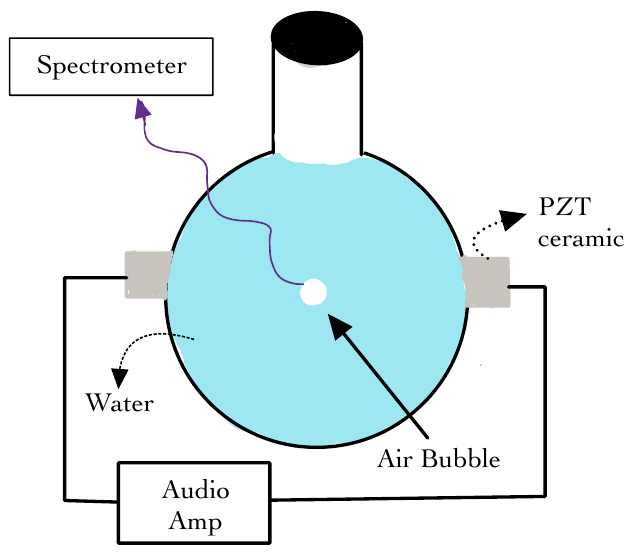}
\caption{A schematic sketch of a typical experimental setup for Sonoluminescence, inspired from the review \cite{Barber, Brenner:2002zz}}\label{sonoexpsetup}
\end{figure}

\begin{figure}[t]
\centering
\includegraphics[scale=0.4]{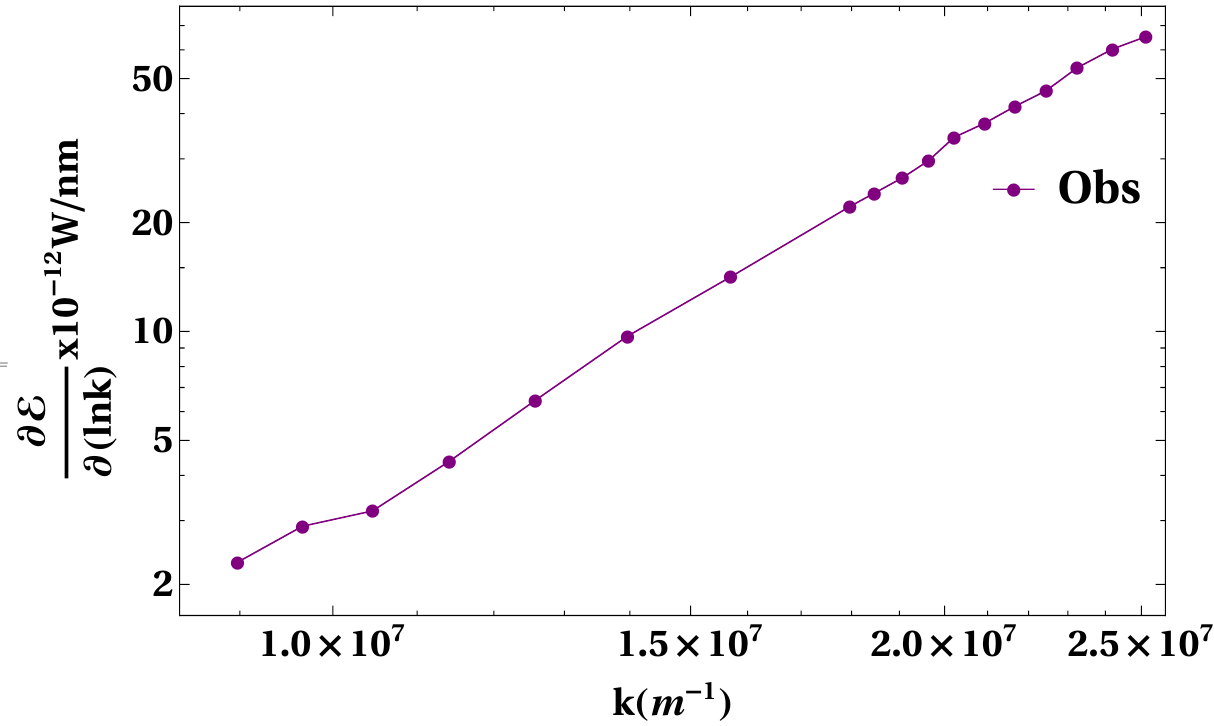}
\caption{A typical Sonoluminescence spectrum has been demonstrated here. The dotted points represent the experimentally measured values which have been extracted from \cite{Hiller:1992qz} and plotted here in k-space.}\label{sonoexpdat}
\end{figure}

It has been experimentally observed \cite{Gaitan, Hiller:1992qz, Gompf, Weninger, Camara} that if an acoustic disturbance acts on a gas bubble in water in such a way that it starts to oscillate quasiperiodically, then this bubble emits repeated flashes of light. Generally, the experimental setup to obtain Sonoluminescence consists of a sealed spherical quartz flask, filled with water, as shown in fig.\ref{sonoexpsetup}. The acoustic drive through the piezoelectric (PZT) ceramic material will induce the perturbation to the air bubble. For stable sonoluminescence, it is found that the required voltage across the PZT is of the order of 50-150 $V$, which will drive the air bubble into resonance with the sound field. Eventually, the bubble will shrink to the minimum radius and the flash of photons will come out.     
The observed flux \cite{Gaitan, Hiller:1992qz, Gompf, Weninger} exhibits a broadband spectrum starting from the visible range to the far ultraviolet range with increasing intensity, as can be seen in the fig.\ref{sonoexpdat}, which consists of data from experimental measurement with acoustic frequency $\sim 27~\SI{}{\kilo\hertz}$. A sudden cutoff appears near $\sim 200~nm$, which matches with the ultraviolet cutoff of water. Sonoluminescence at temperature $22\degree$C fits the black body spectrum which corresponds to the temperature of 25000 K. The efficiency of the light emission heavily depends on the temperature of the water, for example, at $10\degree$ C, the spectrum of sonoluminescence corresponds to black body temperature $\sim$ 50000 K. As already elaborated in the introduction whether the bubble acquired this temperature during the collapse and produced the thermal photon or it is the quantum mechanical phenomena originating from time dependent background is still an unresolved issue \cite{Belgiorno:1999ha}. 
However, our focus will be on the quantum mechanical production. 
\subsection{Dynamics of the bubble}\label{bubble_dynamics}
Experimental measurements suggest the necessary presence, even with a small amount, of an admixture of noble gases such as He, Ar or Xe in the medium inside the bubble. The air medium typically contains a small amount of such inert gases, while the dominant presence of Nitrogen and Oxygen gas will be dissociated during the collapse due to large temperature and subsequently get absorbed in the water through the chemical reactions as has been argued in \cite{Brenner:2002zz}. Thus the bubble, finally, is left with only the inert gas necessary for the sonoluminescence to happen. This made us motivated to consider air as a suitable medium within the bubble as a starting point for our present analysis. However, to this end, we would like to point out that our final result in the present paper does not depend strongly on the characteristics of the medium. In order the take into account the effect of the medium we need to consider photon-medium interaction which we discuss in future.

\begin{figure}[H]
\includegraphics[scale=0.4]{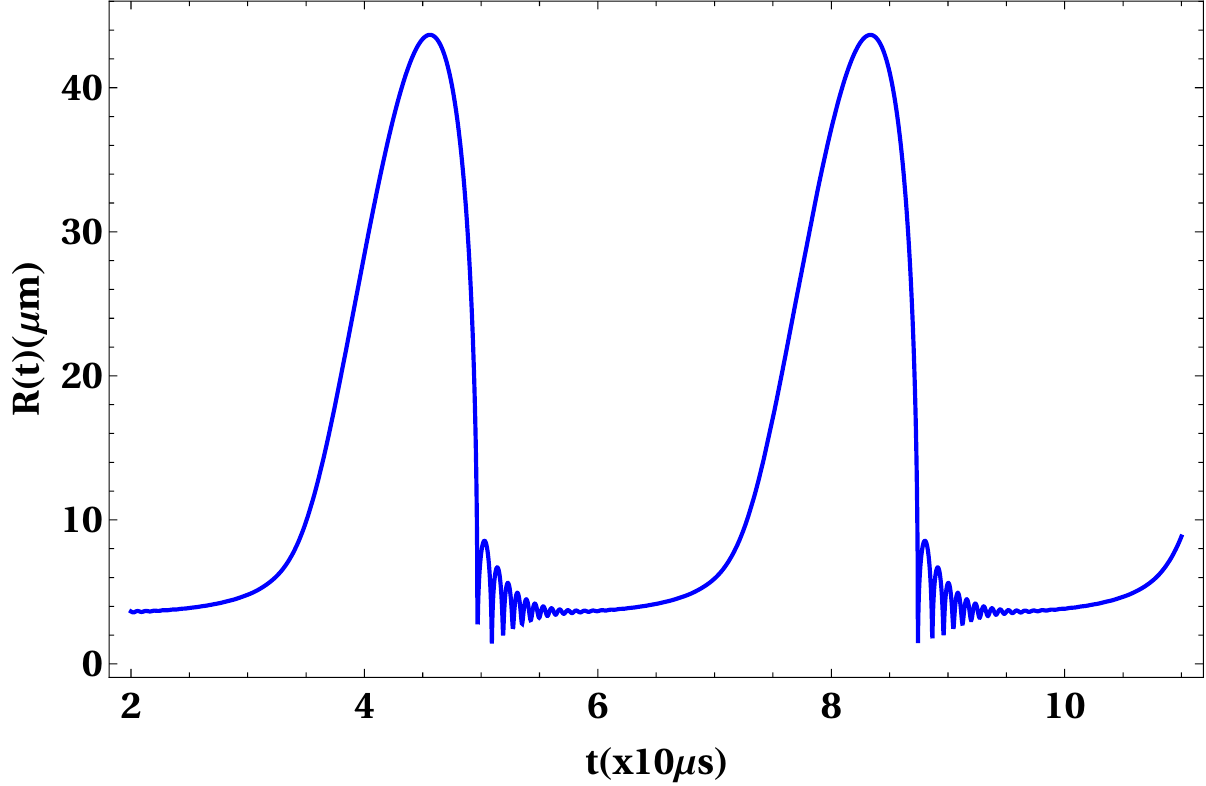}
\caption{Dynamics of the oscillating bubble is described in terms of the temporal profile of the bubble surface. One can see that the RP equation correctly provides for the dynamics of the bubble, which undergoes a repeated quasiperiodic oscillation.}\label{radius}
\end{figure}

The air bubble inside water perturbed under the acoustic wave undergoes rapid oscillation. The dynamics are well captured by the Rayleigh-Plesset (RP) equation \cite{Barber, Rayleigh, Plesset}, arising from the balancing of pressure between the inside and outside medium of the bubble (also see Appendix.\ref{Rp_eqn} for elaborate discussion), and is given as,
\be 
\bea
&R\ddot{R}+\frac{3}{2} \dot{R}^2 \\
&= \frac{1}{\rho} \left( P(R,t) - P_0 + P_a(t) + \frac{R}{c_s} \d{}{t}[P_g(R,t) + P_a(t)]\right), 
\eea
\ee
where, $R(t)$ is the radius of the bubble, $P(R,t)$ is the pressure at the bubble surface, $P_0$ is the constant ambient pressure above the liquid, $P_a(t) = P_a \cos{\omega t}$ is the acoustic drive at the bubble, $P_g(R)$ is the pressure of the gas inside the bubble (as discussed before, we have considered air as the gas medium inside the bubble and corresponding values of the parameters will be given in a moment), $\rho$ is the fluid density, and $c_s$ is the speed of sound in the fluid. The appropriate boundary condition accompanies this above equation,
\be 
P(R, t) + 4 \frac{\eta \dot R}{R} + 2 \frac{\sigma}{R} = P_g(R, t), 
\ee
which arises from the balancing of pressures on either side of the bubble interface. Gas pressure inside the bubble  $P_g(R)$ is,
\be 
P_g(R) = \frac{P_0 R_0^{3 \gamma}}{(R^3 - a^3)^{\gamma}}, 
\ee
where, adiabatic index $\gamma = {C_p}/{C_v} = 1.4$ (for air). Experimental values of other parameters, mentioned above, are as follows: shear viscosity of the fluid, $\eta = 0.003~{\rm Kg/(m-sec)}$, coefficient of the surface tension, $\sigma =0.03~{\rm Kg/sec^2}$, the density of the fluid $\rho\sim 1000~{\rm Kg/m^3}$, pressure-amplitude of the acoustic drive, $P_a = 1.35 $ atm, frequency of the acoustic drive, $\omega_a = 2 \pi (26.5) $ kHz, the minimum radius of the bubble, $R_0 = 4.5 \mbox{ }\mu$m, speed of sound in water, $c_s= 1481$ m/sec, ambient pressure $P_0 = 1$ atm at the minimum radius of the bubble, for an air bubble, van der Waals hard core radius, $a = 0.5~\mbox{}\mu$m. Important to note here that a small change in $\gamma$, $\eta$ and $\sigma$ does not significantly affect the radial dynamics of the bubble as per our numerical analysis (in conformity with the discussion given in Sec. (II.C) of \cite{Brenner:2002zz} ).  Nevertheless, the final form of the RP equation, taking into account the damping of the bubble due to the release of acoustic energy, in terms of the dynamical variable $R(t)$ and the above fluid parameters can be expressed as follows \cite{Barber, Brenner:2002zz},
\be
\bea
&-R \ddot R \left( 1 - \frac{2 \dot R}{c_s} \right) 
-\frac{3}{2} \dot R^2 \left(1-\frac{4}{3}\frac{\dot R}{c_s}\right)\\
&+\frac{1}{\rho}\left[\frac{P_0 R_0^3{\gamma}}{(R^3 - a^3)^{\gamma}} - \frac{4 \eta \dot R}{R} - \frac{2 \sigma}{R} \right]\\
&+\frac{1}{\rho c} \left[\frac{-3\gamma R^3 P_0 R_0^{3 \gamma}}{(R^3 - a^3)^{\gamma + 1}} \dot R+ \frac{2 \sigma}{R} \dot R - \frac{4 \eta}{R} (R \ddot R - \dot R^2)\right]\\
&+ \frac{P_0 R_0^{3 \gamma} \cos{\omega t}}{\rho (R^3 - a^3)^{\gamma}} - \frac{R}{\rho c} P_a \omega \sin{\omega t} - \frac{P_0}{\rho} = 0.
\eea
\ee
We set the initial conditions $R(t=0)=4.5\mu m$ (ambient radius of the bubble) and $R'(t =0) = 0$, although the very nature of the graph does not depend on the initial condition. Using the parameters given above we obtain the dynamics of the bubble radius as plotted in fig.\ref{radius}, 
which happens to agree with experimental measurements \cite{Barber}. It is obvious from fig.\ref{radius} that the bubble undergoes repeated quasi-periodic oscillation. Utilizing this time profile of the bubble dynamics, we will now discuss the analog space time formalism following \cite{Unruh:1980cg}. 

\section{Construction of the Effective Metric}\label{analog_metric} 
In the year 1980, Unruh \cite{Unruh:1980cg} established that the behaviour of the sound on a fluid mimics a field propagating in spacetime background endowed with an effective metric. In this section, we derive this analogue metric sometimes referred to as acoustic metric. For an incompressible and irrotational fluid, the corresponding energy-momentum tensor satisfies the following covariant conservation equation, $\nabla_\mu T^{\mu\nu}=0$,
where,
\be
T^{\mu\nu}=(\rho+P)u^\mu u^\nu+P\eta^{\mu\nu},
\ee
with $\rho$ and $P$ symbolizing density and pressure respectively and $\eta^{\mu\nu}$ is the Minkowski metric describing the flat space time. Whereas, the four-velocity $u^\mu\equiv(1,{\bf v_0})$. Now, as per the consideration, the curl of the velocity being zero leads to the fact that the velocity can be written as ${\bf v}=\nabla \psi_0$, with $\psi_0$ as the velocity-potential for the given fluid. Having the set of fluid equations in hand, we introduce fluctuation as 
\be
\psi\to\psi_0+\bar{\psi}.
\ee
For other fluid parameters, to simplify the computation, we define the following quantity,
\be
g(\zeta)=\int^{e^{\xi}}\frac{1}{\rho'}\frac{dp (\rho')}{d\rho'}d\rho',
\ee
where, $\zeta\equiv\log{\rho}$. Similar to the previous consideration we take the fluctuation on other fluid parameters in terms of $\zeta$ as, $\zeta\to \zeta_0+\bar{\zeta}$. Substituting these perturbed fields in the covariant conservation equation one can derive \cite{Visser:1997ux, Barcelo:2005fc} the fluctuation equation for $\bar{\psi}$ as,
\be\label{eombpsi}
\bea
&\frac{1}{\rho_0} \Bigg[ \pr_t \left( \frac{\rho_0}{g'(\zeta_0)} \right) \pr_t \bar{\psi} + \pr_t \frac{\rho_0 \v v_0}{g'(\zeta_0)}     \cdot \del \bar{\psi} + \del \cdot \left( \frac{\rho_0 \v v_0}{g'(\zeta_0)} \pr_t \bar{\psi}  \right) \\
&- \del \cdot (\rho_0 \del \bar{\psi}) + \del \cdot \left( \v v_0 \frac{\rho_0}{g'(\zeta_0)} \v v_0 \cdot \del \bar{\psi}   \right) \Bigg]=0,
\eea
\ee
which can be expressed in terms of an analog metric (AM) on which $\bar{\psi}$ is dynamical, and the metric is expressed as \cite{Unruh:1980cg},
\be \label{analog}
\bea
&ds_{\rm AM}^2 = \bigg( \frac{\tilde{\rho}_0}{c^2} \bigg)^{-1}\bigg\{  \frac{(\rho_0/c^2)}{(c_0/c)^2}  \bigg[ -\big( \frac{c_0^2}{c^2}  - \frac{v_0^2}{c^2} \big)c^2 \,d{t^2}\\ 
&~~~~~~~~~~~~~~~~~~~~~~~~~~~~~~~~~~~ - 2 \frac{v_i}{c}\,d{x^i} c\,dt  + \delta_{i j} \,d{x^i}\,d{x^j} \bigg]\bigg\}.
\eea
\ee
Here, we have used, $g'(\zeta_0)=g'(\ln \rho_0)\sim c^2_0$, where $c_0$ denotes the local velocity of the sound wave in the medium, e.g., for air $c_0\sim 343~ {\rm m/s}$. Whereas, $\tilde{\rho}_0$ is a new parameter which is introduced to make the $ds_{\rm AM}$ of dimensions length. Moreover, it can also be thought of as a tuning parameter to be adjusted as per the final results. Of course, this overall factor leads to the same fluctuation equation \eqref{eombpsi}. We have two main fluid parameters to be evaluated, the background velocity $v_0$ and the fluid density $\rho_0$. Our interest is to look for the radial behaviour of those parameters for the gas contained inside the bubble. 

Since the system is spherically symmetric, only the radial velocity component will survive, and $v_0$ the velocity of the fluid inside the bubble can be approximately taken as  \cite{Barber}, 
\be
v_0=\frac{\dot R}{R}r ,
\ee 
with physical boundary  conditions $v_0(r = 0) = 0$ and $v_0(r=R)=\dot R$. 
The continuity equation, 
\be
\pd{\rho_0}{t} + \frac{\rho_0}{r^2} \pd{(r^2 v_0)}{r}=0,
\ee
upon using the expression for the background velocity $v_0$, yields,
\be
\rho_0=\rho_{eq}\frac{R^3_0}{R^3(t)}
\ee 
where, $\rho_{eq}=\rho_0(t\to t_0)$, at the minimum radius of the bubble say  $R_0=R(t\to t_0)$. The above expression is consistent with the ideal gas equation  $\rho_0=\frac{P_0 R_0^3}{c_0^2}\frac{1}{R^3}$, where, as it turns out, $P_0 =\rho_{eq}c^2_0$.
Therefore, we rewrite the factor involving density as $\frac{\tilde{\rho}_0^{-1}\rho_0}{c^2_0}= \frac{\xi^3}{R(t)^3}$, where $\xi$ is the new arbitrary parameter, incorporating the old ones.

Our goal is to utilize the analog geometry derived in Eq.\ref{analog} to couple with the EM field. In all the previously discussed quantum models, to the best of our knowledge analog geometric approach has not been taken into account. It is important to note that electromagnetism in four dimensions is conformal invariant. Despite non-trivial time dependence any conformal modification to the background, therefore, should not lead to any physical production of photons from the quantum vacuum. One such modification that has been discussed is to introduce the time dependent dielectric constant $(\epsilon) $\cite{Liberati:1998wg, Visser:1998bqu, Liberati:1999jq, Liberati:1999uw} through refractive index. Typically such modification appears in the time part of the spacetime which is of conformal type. Naive application of such an approach usually leads to divergent results \cite{Milton:1996wm, Milton:1997ky}. Such divergence can be done away with by conformal transformation in the background.

In our present submission, we conjecture that the EM field perceives such analog metric as a fluctuation on the usual flat spacetime geometry as follows,
\be
\bea
ds^2&=\bigg( - \frac{\,d{t^2}}{\epsilon} + \,d{r^2} + r^2 \,d{\Omega^2} \bigg)\\
&+  \frac{\xi^3}{R(t)^3}\bigg[ -(c_0^2 - v_0^2) \,d{t^2} - 2v_0drdt+d{r^2}+r^2d{\Omega^2}\bigg]\\
&= (g_{\mu\nu}^{(0)} + h_{\mu \nu}) \,d{x^{\mu}} \,d{x^{\nu}}.
\eea
\ee
 Note that we have used the natural unit, $c=1$, which will be followed in the later sections unless otherwise stated.
Here, $\epsilon\sim 1$ represents the dielectric constant of the air medium, as the above metric effectively describes the medium inside the bubble. Hence, there is no significant change in the dielectric constant. We argue in the later sections that it's not the dynamical boundary separating two different mediums, but the quasi-oscillation of the fluid parameters inside the bubble that is solely responsible for Sonoluminescence. {\it In this respect, we significantly differ from the existing formalisms\cite{schwinger1992a,schwinger1992b,schwinger1993a,schwinger1993b,schwinger1993c, schwinger1993d, schwinger1994a, Liberati:1998wg, Visser:1998bqu, Liberati:1999jq, Liberati:1999uw, Eberlein:1995ex, Eberlein:1995ev} of Sonoluminescence based on the dynamical Casimir effect.}

We consider the inverse of the fluctuation metric as, $h^{\mu \nu} = g^{\mu \alpha}_0 g^{\nu \beta}_0 h_{\alpha \beta}$, so that, any raising and lowering of indices should be with respect to $g^{(0)}_{\mu\nu}$. 
The total metric can be expressed as 
\be
ds^2=-f(t,r)dt^2+2g(t,r)drdt+p(t)\left(dr^2+r^2d\Omega^2\right),
\ee
with the essential matrix elements taking the following form,
\be\label{offdiag}
\bea
&p(t) = 1 + \frac{\xi^3}{R^3},\\
&f(t, r) =1 + \frac{\xi^3}{R^3} \left( c_0^2 - \frac{\dot{R}^2}{R^2} r^2 \right), \\
&g(t, r) = - \frac{\dot{R} \xi^3}{R^4} r.
\eea
\ee
To simplify our task we can diagonalize the above metric making use of the following transformation for the radial coordinate to bring it to a more convenient form, 
\be
\frac{\,d \bar{r}}{{1}/{p^{1/6}}} = \sqrt{p} \,d r + \frac{g}{\sqrt{p}} \,dt,
\ee
which translates to the rescaling of the radial coordinate, one can derive, $\bar{r} = r p^{1/3}$ (for details of this derivation, see appendix \ref{diagonal}). Using this redefinition, the metric reduces to, 
\be\label{scaled_metric}
\,d s^2 = - \left( f + \frac{g^2}{p} \right) \,dt^2 + p^{1/3} \left( \,d \bar{r}^2 + \bar{r}^2 \,d{\Omega}^2 \right).
\ee
In our subsequent analysis, we will just replace the bar sign in the radial coordinate and treat it as usual, $r$. As argued before the photon production essentially occurs due to the dynamics of the fluid medium inside the bubble. Furthermore, experimentally the photon flux is observed to originate when the oscillating bubble is nearly at its minimum radius \cite{Gaitan, Barber, Brujan}. Inspired by this observation, we impose the $r\to 0$ limit throughout our analysis. Thanks to this limit which leads the fluctuation part ($h_{\mu\nu}$) transformed into a simplified form, so that the metric coefficients \eqref{offdiag} become,
\be\label{coeff}
\bea
& p(t)=1+\frac{\xi^3}{R^3},\\
&f(t)=1+\frac{\xi^3}{R^3}c_0^2\\
& g=0.
\eea
\ee
In the following discussion, we will consider the metric \eqref{scaled_metric} with the above metric coefficients. 
\begin{figure}[t]
\includegraphics[scale=0.4]{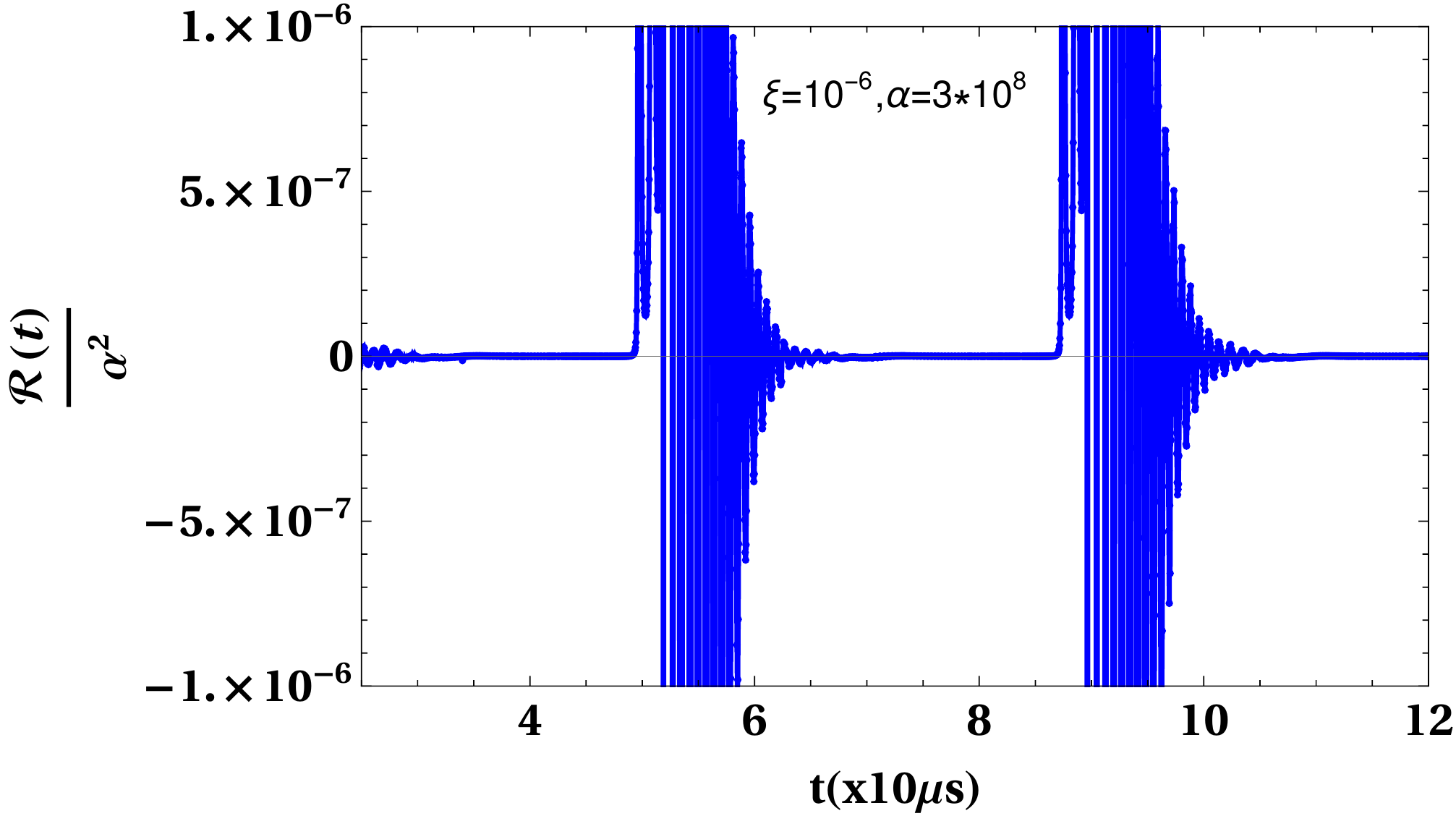}
\caption{Evolution of the curvature scalar of the analog metric, with time }\label{ricciplot}
\end{figure}
\section{General formalism}
Simply keeping the Lorentz invariance intact, generically, one can couple the EM field to the background space time in the linear order in curvature as shown below \cite{Balakin:2005fu, Hehl:1999bt},
\be\label{gen.lagrang}
\bea
{\cal L}=&-\frac{\sqrt{-g}}{4}\Big[F_{{\mu}{\nu}}F^{\mu\nu} +\chi^{\mu\nu\alpha\beta}(q_1 F_{{\mu}{\nu}}F_{\alpha\beta}+q_2 F_{{\mu}{\nu}}\tilde{F}_{\alpha\beta})\Big],
\eea
\ee
with
\be
\bea
\chi^{\mu\nu\alpha\beta}&=\alpha_1\mathcal{R}(g^{\mu\alpha}g^{\nu\beta}-g^{\mu\beta}g^{\nu\alpha})\\
&~~~+\alpha_2(R^{\mu\alpha}g^{\nu\beta}-R^{\mu\beta}g^{\nu\alpha}+R^{\nu\beta}g^{\mu\alpha}-R^{\nu\alpha}g^{\mu\beta})\\
&~~~+\alpha_3R^{\mu\nu\alpha\beta},
\eea
\ee
where $F_{\mu\nu}$ is the field strength tensor of the EM field and $\tilde{F}^{\mu\nu}=\epsilon^{\mu\nu\alpha\beta}F_{\alpha\beta}/2$ is the dual tensor. Whereas, $(q_1, q_2)$ and $(\alpha_1, \alpha_2, \alpha_3)$ are arbitrary constants need to be fixed.
Here, $\mathcal{R}, R^{\mu\nu}$ and $R^{\mu\nu\alpha\beta}$ represent the Ricci scalar, Ricci tensor and Riemann tensor of the background metric
\eqref{scaled_metric}\eqref{coeff} respectively. Now the metric \eqref{scaled_metric} with \eqref{coeff} can be recast into a conformally flat form as,
\be
\,d s^2 =  p(\tau)^{\frac 1 3}(-d\tau^2 + dx^2 + dy^2 + dz^2) ,
\ee
with conformal time $d\tau = dt\sqrt{f/p^{1/3}}$. Keeping the symmetry intact, we have used Cartesian coordinate to describe the spatial part of the metric to simplify the analysis. Throughout our analysis, we will work assuming the parameter $\xi = 10^{-6}$ (given in unit of $R$), such that $\xi/R(t) < 1$, which compels us to consider $\tau\simeq t$, i.e. conformal time approximated as real time. Important to note that having this conformally flat metric, the first term in the given Lagrangian does not lead to any real particle production as the minimally coupled action of the EM field in this space time is conformally invariant \cite{Codirla:1997mz, Cote:2019kbg}. The rest of the terms in the Lagrangian represent non-minimal coupling and lead to particle production. Furthermore, the coupling with the dual tensor ($F\tilde{F}$) signifies a parity breaking term that leads to the production of the helical EM field (one may look at  \cite{Tripathy:2021sfb, Tripathy:2022iev, Talebian:2021dfq, Adshead:2016iae}, where such considerations are often studied in the context of magnetogenesis in early universe cosmology). We do not consider such terms leading to non zero helicity which may interesting in the context of observation, and we will consider this in our future course of study. Therefore we will put $q_2=0$ for the present work. The coupling with Reimann tensor and Ricci tensor, however, is equally important as the Ricci coupling with the $FF$. For the simplification of the numerical computation, we restrict ourselves to the Ricci coupling only, for the present analysis, keeping $\alpha_2=\alpha_3=0$ (otherwise, the equation of motion of the EM field will involve 4 non-zero components of the Ricci tensor and 12 non-zero components of the Riemann tensor for the metric under consideration). The analysis with consideration of these other couplings, we leave for our future work.

\subsection{Quantization of the EM field}\label{quantize} 
We consider the following Lagrangian for the Maxwell field with conformal symmetry breaking coupling,
\be\label{action}
{\cal L}=-\frac{\sqrt{-g}}{4} \left[F_{{\mu}{\nu}}F^{\mu\nu}+\frac{\mathcal{R}(t)}{\alpha^2}F_{{\mu}{\nu}}F^{\mu\nu}\right],
\ee
where $\alpha^2\equiv 1/(\alpha_1q_1)$ from \eqref{gen.lagrang}.
The Ricci scalar of the background metric turns out to be,
\be
\bea
\mathcal{R}(t)\simeq-3\frac{\xi^3}{R^3}\left[\frac{\pr^2_t R}{R}-4\left(\frac{\pr_t R}{R}\right)^2\right],
\eea
\ee
which captures the oscillating features of the bubble, as can be seen from fig.\ref{ricciplot}. $\alpha$ is a controlling parameter of mass dimension unity.  We ignore all the higher order $\xi^n$, $n>3$ coupling. We symbolize the conformal breaking coupling factor as $\mathcal{I}(t)=1+{\mathcal{R}(t)}/{\alpha^2}$. 
We should also mention that breaking the conformal invariance, considering such a coupling prescription is not totally new, as one may find in \cite{Tripathy:2021sfb, Tripathy:2022iev, Talebian:2021dfq, Adshead:2016iae}, and the formulations of non-perturbative production in curved space time are also well studied \cite{Kofman:1997yn}. However, to the best of our knowledge, we have not found any work where such analyses have been performed in the context of analog systems.   

Throughout our analysis, we fix $\alpha = 3\times 10^{8}$ in the unit of Ricci scalar, $\mathcal{R}$, such that $|\mathcal{I}(t)-1|<1$.
With this setup, one obtains the equation of motion by varying the action as, 
\be\label{eom1}
\pr_\mu\Big(\sqrt{-g}g^{\mu\alpha}g^{\nu\beta}\mathcal{I}(t)F_{\alpha\beta}\Big)=0.
\ee
The spatial part of the metric being flat, we can suitably choose the Coulomb gauge condition,
\be
A_t=0, ~~ \nabla\cdot{\bf A}=0.
\ee

Only with time-dependent coupling, we can expand the vector potential in Fourier mode in the following manner,
\be\label{cfield}
A_i(t,{\xb})=\int\frac{d^3k}{(2\pi)^3}[c_{\kb}A_i(t,k)e^{\mi{\kb}\cdot{\xb}}+c^*_{\kb}A^*_i(t,k)e^{-\mi{\kb}\cdot{\xb}}],
\ee
where $c_{\kb}$, and $c^*_{\kb}$ are some arbitrary constants for the time being.
Substituting the above mode expansion of the EM field in \eqref{eom1} we arrive at, 
\be\label{eommode}
\pr^2_t A_i(t,k)+\frac{\pr_t\mathcal{I}(t)}{\mathcal{I}(t)}\pr_t A_i(t,k)+|{\bf k}|^2A_i(t, k)=0,
\ee
where we have used $g^{nl}\pr_n A_l=0$, i.e. the gauge condition and $k^2=|{\kb}|^2=\delta^{nl}k_nk_l$ denotes the square of the amplitude of wave vector, ${\bf k}$. In the following discussion, we demonstrate the quantization procedure of this classical field. 

\subsection{Calculatoin of Bolgoliuobov coefficients}
We now write the classical EM field \eqref{cfield} in polarization basis and promote it to the quantum field operator by making the constants, $c_{\kb},c^*_{\kb}$, as creation and annihilation operator respectively,
 \be\label{qmfield}
\bea
&A_i(t,{\xb})\\
&=\sum_{\lambda=1,2}\int\frac{d^3k}{(2\pi)^3}\epsilon^\lambda_i({\kb})[\hat{c}^\lambda_{\kb}A_\lambda(t,k)+\hat{c}^{\lambda\dagger}_{-\kb}A^*_\lambda(t,k)]e^{-\mi{\kb}\cdot{\xb}},
\eea
\ee
where, $\hat{c}^\lambda_{\kb}$ and $\hat{c}^{\lambda\dagger}_{\kb}$ satisfy the usual harmonic oscillator commutation relation, 
\be
\bea
&[\hat{c}^\lambda_{\kb},\hat{c}^{\lambda'}_{\kb'}]=0,~~[\hat{c}^{\lambda\dagger}_{\kb},\hat{c}^{\lambda'\dagger}_{\kb'}]=0,\\
&[\hat{c}^\lambda_{\kb},\hat{c}^{\lambda'\dagger}_{\kb'}]=(2\pi)^3\delta^{\lambda\lambda'}\delta^3({\kb-\kb'}).
\eea
\ee
Whereas orthonormalized polarization basis vectors satisfy the following relation
\be
\bea
&\epsilon^\lambda_i({\kb})\cdot k_i=0,~~ \epsilon^\lambda_i({\kb})\epsilon^{\lambda'}_i({\kb})=\delta_{\lambda\lambda'},\\ &\sum_{\lambda=1,2}\epsilon^\lambda_i({\kb})\epsilon_\lambda^j({\kb})=\delta_i^j-\frac{k_ik^j}{k^2}.
\eea
\ee
From the action \eqref{action} the canonically conjugate momentum of $A_i(t,\v x)$ turns out to be,
\be
\Pi^i=-\sqrt{-g}\mathcal{I}(t)g^{00}g^{ij}\pr_\eta A_j=\mathcal{I}(t)\delta^{ij}\pr_t A_j.
\ee
Imposing the equal time canonical commutation relation between the EM potentials and their conjugate momentum we obtain
\be\label{com}
[A_i(t,{\xb}),\Pi^j(t,{\yb})]=\mi\int\frac{d^3k}{(2\pi)^3}e^{\mi{\kb}\cdot{(\xb-\yb)}}\epsilon^\lambda_i({\kb})\epsilon^j_\lambda({\kb}).
\ee
where, $\epsilon^\lambda_i({\kb})\epsilon^j_\lambda({\kb})=\Big(\delta_i^j-\frac{k_ik^j}{k^2}\Big)$. From this commutation relation we can deduce the quantization condition for the EM potentials as \cite{Bamba:2003av},
\be\label{qunt.cond}
A_\lambda(t,k)A'^*_\lambda(t,k)-A^*_\lambda(t,k)A'_\lambda(t,k)=\frac{i}{\mathcal{I}(t)},
\ee
where, $A'_\lambda(t,k)=\pr_t A_\lambda(t,k)$ also $\delta^{\ij}\epsilon^\lambda_i({\kb})\epsilon^{\lambda'}_j(-{\kb})=\delta_{\lambda\lambda'}$.
We assume each mode starts its journey from the quantum vacuum with no particle state at an initial time, $\hat{c}^{\lambda'\dagger}_{\kb'}|0\rangle = 0$, which same as Bunch-Davies\cite{Birrell:1982ix, Allen:1985ux} vacuum considered in cosmology. The associated mode function takes the form, $A_\lambda(t,k)\sim N_ke^{-ikt}$, with positive frequency outgoing mode, and the initial normalization can be fixed using Eq.\ref{qunt.cond}, as $N_k=1/\sqrt{2k\mathcal{I}(t_0)}$.
Let's write the Fourier mode terms of \eqref{qmfield} as
\be
\bea
&\hat{B}_\lambda(t,{\kb})=\hat{c}^\lambda_{\kb}A_\lambda(t,k)+\hat{c}^{\lambda\dagger}_{-\kb}A^*_\lambda(t,k),\\
&\hat{\pi}_\lambda(t,{\kb})=\mathcal{I}(t)[\hat{c}^\lambda_{\kb}\dot{A}_\lambda(t,k)+\hat{c}^{\lambda\dagger}_{-\kb}\dot{A}^*_\lambda(t,k)].
\eea
\ee
Introducing another set of time dependent creation and annihilation operators in terms of the above field operators as, 
\be\label{latetimeop}
\bea
\hat{d}^\lambda_{\kb}(t)&\equiv \sqrt{\frac{\mathcal{I}(t)k}{2}}\hat{B}_\lambda(t,{\kb})+\frac{\mi}{\sqrt{2\mathcal{I}(t)k}}\hat{\pi}_\lambda(t,{\kb}),\\
\hat{d}^{\lambda\dagger}_{\kb}(t)&\equiv \sqrt{\frac{\mathcal{I}(t)k}{2}}\hat{B}_\lambda(t,-{\kb})-\frac{\mi}{\sqrt{2\mathcal{I}(t)k}}\hat{\pi}_\lambda(t,-{\kb}),
\eea
\ee
we define late time creation and annihilation operators expressed in terms of respective initial time operators and time dependent Bogoliubov coefficients \cite{Birrell:1982ix} in the following manner:
\be\label{bogcoef}
\bea
&\hat{d}^\lambda_{\kb}(t)=\alpha^\lambda_k(t)\hat{c}^\lambda_{\kb}+\beta^{\lambda^*}_k(t)\hat{c}^{\lambda\dagger}_{-\kb},\\
&\hat{d}^{\lambda\dagger}_{\kb}(t)=\alpha^{\lambda *}_k(t)\hat{c}^{\lambda\dagger}_{\kb}+\beta^\lambda_k(t)\hat{c}^{\lambda}_{-\kb}.
\eea
\ee
The time dependent Bogoliubov coefficients are given by \cite{Kobayashi:2019uqs, Haque:2020bip},
\be\label{Bog}
\bea
&\alpha^\lambda_k(t)=\sqrt{\mathcal{I}(t)}\Big(\sqrt{\frac{k}{2}}A_\lambda(t,k)+\frac{\mi}{\sqrt{2k}}A'_\lambda(t,k)\Big),\\
&\beta^\lambda_k(t)=\sqrt{\mathcal{I}(t)}\Big(\sqrt{\frac{k}{2}}A_\lambda(t,k)-\frac{\mi}{\sqrt{2k}}A'_\lambda(t,k)\Big).
\eea
\ee
From the quantization condition \eqref{qunt.cond} one can derive the following constrained relation of the Bogoliubov coefficients,
\be
|\alpha^\lambda_k(t)|^2-|\beta^\lambda_k(t)|^2=1.
\ee
Now using the expression of $\beta^\lambda_k(t)$ \eqref{Bog} and with the relation \eqref{qunt.cond} we obtain
\be\label{betksqr}
\bea
|\beta^\lambda_k(t)|^2=\frac{\mathcal{I}(t)}{2}\Big(k|A_\lambda(t,k)|^2+\frac{|A'_\lambda(t,k)|^2}{k}\Big)-\frac{1}{2},
\eea
\ee
which physically represents the photon number density produced from the quantum vacuum due to time dependent background \cite{Mukhanov:2007zz, Domcke:2021fee, Ema:2018ucl}. 

 In the following discussion, we will analyze the time evolution of this quantity and present the interpretation of the produced photon flash from the growth of $|\beta^\lambda_k(t)|^2$. In our framework, we have introduced two arbitrary parameters $\alpha$ and $\xi$. Depending upon those parameter values, we show particles can be produced both through perturbative and non-perturbative processes. However, it is generically true and also we show that for the perturbative process, the particle production is insignificant, and it is the non-perturbative resonance due to the oscillating bubble background which will contribute the most. To study this analogy, we proceed by numerically finding out the solution of the EM field from the equation of motion \eqref{eommode}. It is important to note that the extreme stiffness of the Ricci scalar and its time derivative, make this equation very stiff. When dealing with stiff ordinary differential equations, it is well known that implicit numerical methods, characterized by a greater number of discretization steps, work better than explicit methods in ensuring stable solutions \cite{Teukolsky:1986, KOCH2000231}. Naturally, this advantage comes at the cost of increased time consumption. For our present analysis, we have used such an implicit, Cranck-Nicholson method\cite{Bao, Adhikari} (see also Appendix.\ref{CN}).
 \begin{figure}[t]
\includegraphics[scale=0.4]{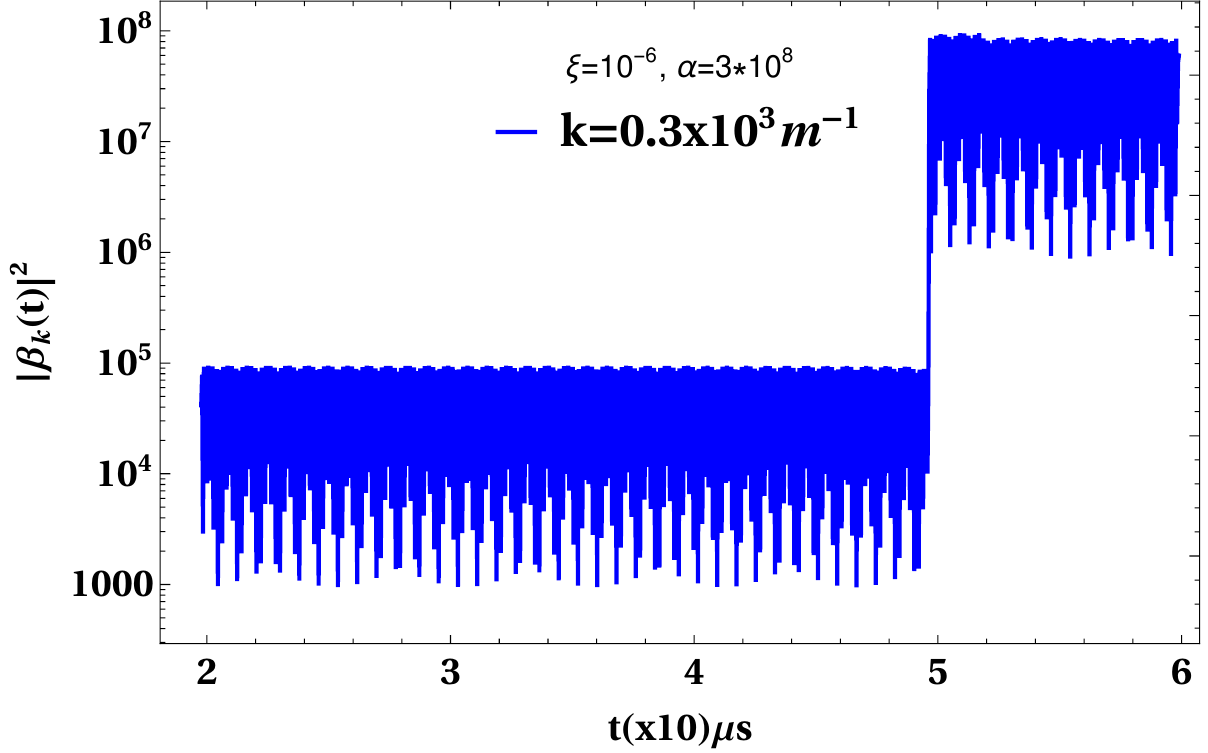}
\caption{Spectral number density has been plotted with time for a fixed $k=0.3\times 10^3~ m^{-1}$. In the Y-axis we have used a log scale.}\label{betaksqrwt}
\end{figure} 

Using the solution of the EM field in the expression  \eqref{betksqr} we have plotted $|\beta^\lambda_k(t)|^2$ with time in fig.\ref{betaksqrwt} for a fixed frequency. It is indeed observed that at the moment say $t=50 {\rm \mu s}$ when the bubble comes to its minimum radius (fig.\ref{radius}) a sudden enhancement in the photon number density appears. Most importantly, the enhancement gets repeated with the time period of the bubble shrinking to its minimum radius. This indeed indicates parametric resonance due to the breakdown of adiabaticity. We interpret this enhancement of the energy density at each successive period as the indicator of the periodically emitted photon flux as observed in the experiment \cite{Gaitan, Hiller:1992qz, Gompf}. The sudden growth of number density at the time of collapse is expected to occur within a specific frequency band which is a typical feature of the Fluquet system \cite{Lozanov:2019jxc}, and this is indeed a significant result of our present proposal accounting for the experimental findings\cite{Hiller:1992qz}. 
The particle production happens at the moment when the bubble comes to its minimum radius. Therefore, by tuning $(\alpha,\xi)$ values where the effect is maximized, surprisingly we obtain the magnitude of the number spectrum $|\beta_k|^2$ very close to that of the experimentally observed value, as we will see in a moment. 

\begin{figure}[t]
\includegraphics[scale=0.4]{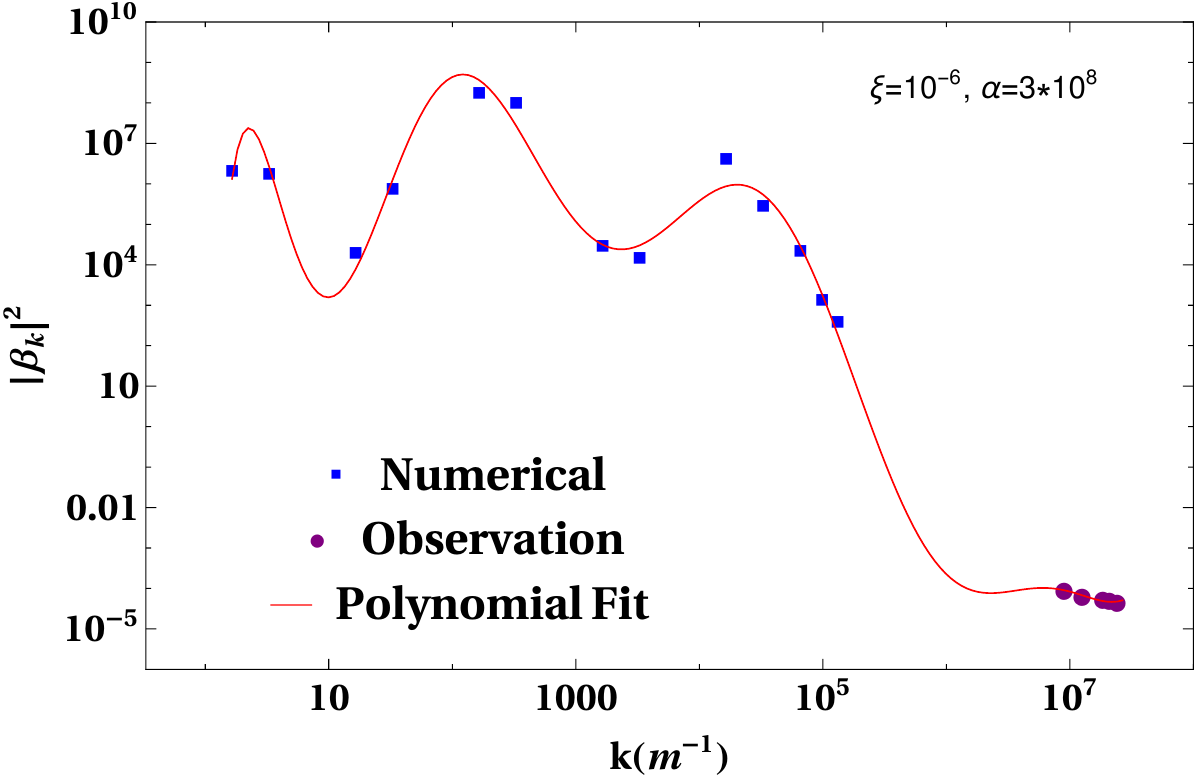}
\caption{The spectral number density is plotted with frequency using a log scale for both axes. The experimental data is given along with the data numerically calculated from our model for comparison. We have fitted both results using Polynomial fit in Mathematica.}\label{betaksqrwk}
\end{figure}

Under our adopted numerical method, we could get the spectrum in the range $(1, 10^5)~\mbox{m}^{-1}$ 
of $k$, as shown in blue dotted points fig.\ref{betaksqrwk}. 
On the same plot we also provide data for the experimentally observed frequency ranges \cite{Hiller:1992qz} around $10^7~\mbox{m}^{-1}$ that extend from visible to ultraviolet range reviling the broadband nature of the spectrum, and is shown in purple dots. As previously stated, the implicit numerical method might be the most effective approach. However, even following such a numerical technique, the Crank-Nicholson method, we could not get a reliable solution for $k >10^5~\mbox{m}^{-1}$. Nevertheless, the trend of the numerically obtained photon number appears to indicate that our analog metric framework could explain the actual observation. This motivated us to extrapolate our results taking into account the experimentally measured data using the Polynomial fit (look at Appendix.\ref{fitfunc} for the fitting function), which is highlighted in red coloured line in fig.\ref{betaksqrwk}. The nice fitting of the extrapolated curve accounting for the experimental data seems to suggest that vacuum production through parametric resonance could be a possible mechanism of sonoluminescence. To this end, we should mention that addressing the relevant numerical challenges requires a thorough investigation into more robust numerical analysis techniques, such as implementations of the implicit methods discussed in \cite{Elrod:2022, Hairer:1999, Hairer:1996}. In particular, the Parallelization technique presented in \cite{Elrod:2022}, aimed at addressing time consumption in the implicit method, could be a suitable approach. By employing these methods, it may become possible to address the stiffness problem in the model equations, thereby obtaining a stable solution within the experimentally observed frequency range. Then, it will be possible to confirm whether the current formulations are capable of accurately reproducing the fitted line in the high-frequency range, aligning with the actual measured data, or if they might deviate from it.
In the next section, we calculate photon energy flux that has also been experimentally measured.  
\subsection{Photon Energy Flux}\label{energyflux}
\begin{figure}[t]
\includegraphics[scale=0.4]{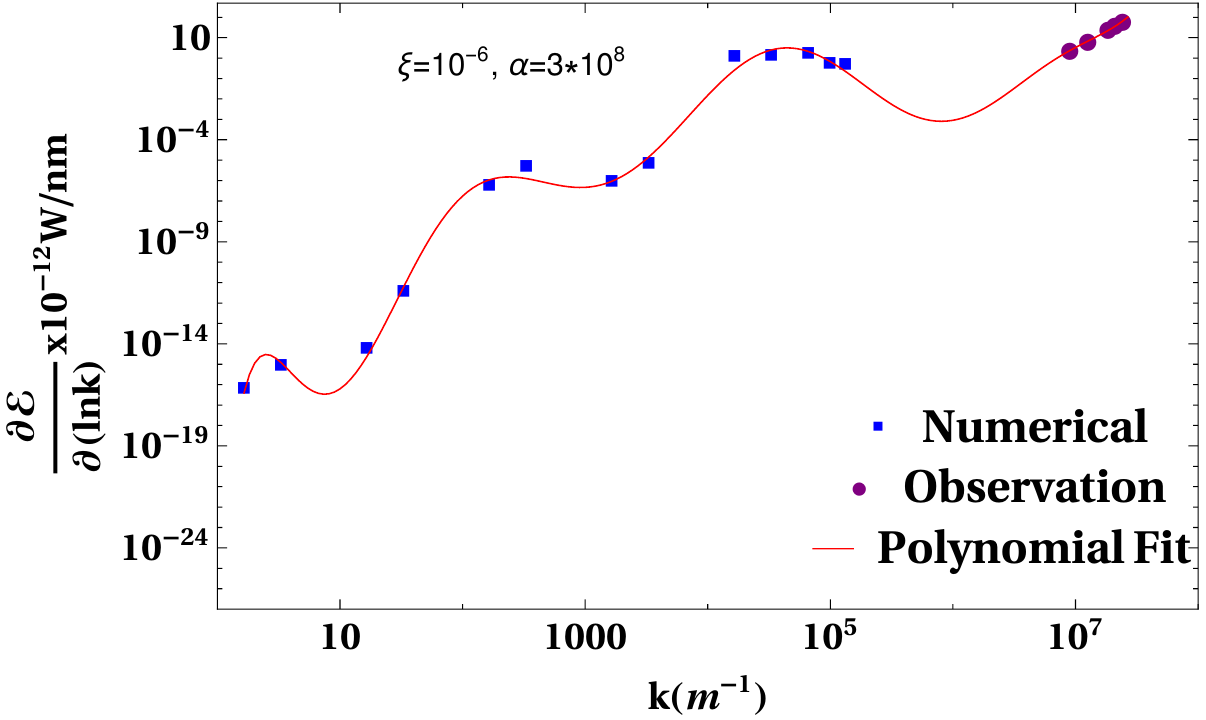}
\caption{The power spectrum of the photon emission from the sono bubble is demonstrated using a log scale for both axes. The experimental data is given along with the data numerically calculated from our model for comparison. We have fitted both results using Polynomial fit in Mathematica.}\label{flux}
\end{figure}

To connect with the experimental observation we evaluate the spectral energy density (energy density per $k$ mode) from the Hamiltonian that can be calculated from the Lagrangian \eqref{action}. The final expression for Hamiltonian in terms of creation and annihilation operators (with the help of \eqref{latetimeop}) can be expressed as,
\be
H=\sum_{\lambda=1,2}\int\frac{d^3k}{(2\pi)^3}~k\hat{d}^{\lambda\dagger}_{{\kb}}(t)\hat{d}^{\lambda}_{\kb}(t).
\ee
The expectation value of this Hamiltonian operator on the initial vacuum state discussed in Sec.\ref{quantize}, turns out as,
\be
\langle H\rangle=\sum_{\lambda=1,2}\int\frac{d^3k}{(2\pi)^3}~k|\beta^\lambda_k(t)|^2\delta^3(0),
\ee
where $\delta^3(0) = \mbox{Volume}$ in the above equation appears from the commutators of the creation and annihilation operators, signifies infinite spatial volume emerging due to the quantization of the EM field throughout the space-time \cite{Mukhanov:2007zz}. Therefore, the energy density comes out to be,
\be
\mathcal{E}=\frac{\langle H\rangle}{\delta^3(0)} =\sum_{\lambda=1,2}\int dk~4\pi k^3|\beta^\lambda_k(t)|^2,
\ee
and the associated spectral energy density assumes the following form,
\be
\frac{\pr\mathcal{E}}{\pr\ln k}=\sum_{\lambda=1,2}~4\pi k^4|\beta^\lambda_k(t)|^2.
\ee
The dimension of the above quantity is simply energy per unit volume. Converting this into the unit of  experimentally measured quantity as 
\be
\frac{\mbox{Flux}}{s \times \mbox{nm}} = \frac{\pr\mathcal{E}}{\pr\ln k} \times \frac{4\pi(0.2\mu\mbox{m})^2}{50{\rm ps}} \left[\frac{\mbox{Watt}}{\mbox{nm}}\right],
\ee
where, the typical radius of the bubble at the time of the emission is considered as $\sim 200 nm$\cite{Camara}, and we have used experimentally measured pulse width $\sim 50{\rm ps}$ \cite{Hiller:1992qz, Weninger}.
we numerically evaluate the above expression using  \eqref{betksqr}, and plotted in Fig.\ref{flux}.  Interestingly, one can see the increasing trend in the flux, shown in blue dotted points, with the wave number $k$. The nature of the spectrum indicates that it sufficiently produces the required amount of flux which is close to the observed one, shown in purple dotted points. For example, we obtained the maximum numerical value of flux, $1.5\times 10^{-14}$Watt/nm for a frequency $\sim 0.7\times 10^5 {\rm m}^{-1}$.
The upward trend of the flux seems to suggest that by invoking an improved numerical method one can indeed reach the level of experimentally observed amplitude $\sim 5\times 10^{-12}$ Watt/nm near around $k \sim 10^7 {\rm m}^{-1}$ since the production is mainly driven by parametric resonance. It might also happen that in the observed frequency range, our model could not reach the measured amplitude of the photon flux (the same issue discussed in the previous subsection with the possible way out). Whether the nature of our results can have the potential to match the experimentally observed flux, we have extrapolated our results including measured data points, utilizing the polynomial fit used in fig.\ref{betaksqrwk}, and highlighted the fitting function with a red solid line in fig.\ref{flux}. 

\section{Perturbative spectrum at high-Frequency: analytical Estimation}\label{highfreq_flux}
In this section, we discuss perturbative photon production in the high frequency limit. We use the well known WKB method to get the analytic spectrum and also show the emergence of divergent contributions if the conformal property of the EM field is not appropriately taken into account. To use the WKB method we rewrite the EM mode Eq. \eqref{eommode} in terms of newly defined field, $\tilde{A}_i(t,k)=\sqrt{\mathcal{I}(t)}A_i(t,k)$, and obtain the modified field equation as,
\be\label{tildeeom}
\pr^2_t\tilde{A}_i(t,k)+\omega^2_k(t)\tilde{A}_i(t,k)=0,
\ee
where the time dependent frequency is expressed as
\begin{equation}
\omega^2_k(t)=\frac{\dot{\mathcal{I}}^2(t)}{4\mathcal{I}^2(t)}-\frac{\dot{\mathcal{I}}(t)}{2\mathcal{I}(t)}+\epsilon(t) k^2\equiv \mathcal{J}(t)+\epsilon(t) k^2 ,
\end{equation}
with $\mathcal{J}(t)\equiv \frac{\dot{\mathcal{I}}^2(t)}{4\mathcal{I}^2(t)}-\frac{\ddot{\mathcal{I}}(t)}{2\mathcal{I}(t)}$. While the factor $\epsilon(t)$ multiplied with $k$ is the time dependent dielectric constant of the fluid inside the oscillating bubble. In our previous analysis, we assumed it to be unity. This particular time dependent term can be identified with the conformal modification of the flat Minkowski background, as it can always be absorbed into the modified time. However, without any time modification, we argue that the presence of such a term will always lead to divergent contributions indicating the non-physical contribution to the photon flux. 

Remembering the well known WKB method in the high frequency limit, the temporal part of the EM field $\tilde{A}_i(t,\textbf{k})$ is expressed as\cite{Ema:2019yrd},
\be
\bea
&\tilde{A}_\lambda(t,k)=\frac{\gamma^\lambda_k(t)}{\sqrt{2\omega_k(t)}}e^{-\mi\int\omega_k(t)~dt}+\frac{\rho^\lambda_k(t)}{\sqrt{2\omega_k(t)}}e^{\mi\int\omega_k(t)~dt}.\\
\eea
\ee
Utilizing this the original EM field mode assumes,  
\be
\bea
&A_\lambda(t,k)=\frac{\gamma^\lambda_k(t)e^{-\mi\int\omega_k(t)~dt}}{\sqrt{2\mathcal{I}(t)\omega_k(t)}}+\frac{\rho^\lambda_k(t)e^{\mi\int\omega_k(t)~dt}}{\sqrt{2\mathcal{I}(t)\omega_k(t)}}.\\
\eea
\ee
Comparing the above expression with \eqref{Bog} we deduce the expression for the Bogoliubov coefficients as, 
\be
\bea
&\alpha^\lambda_k(t)=\gamma^\lambda_k(t)\sqrt{\frac{k}{\omega_k(t)}}e^{-\mi\int\omega_k(t)~dt},\\
&\beta^\lambda_k(t)=\rho^\lambda_k(t)\sqrt{\frac{k}{\omega_k(t)}}e^{\mi\int\omega_k(t)~dt}.
\eea
\ee
Finally, the EM field equation can be equivalently written in terms of WKB functions from \eqref{tildeeom} as the following first order differential equations \cite{Ema:2019yrd},
\be
\bea
&\dot{\gamma}^\lambda_k(t)=\frac{\dot{\omega}_k(t)}{2\omega_k(t)}e^{2\mi\int^t\omega_k(t')dt'}\rho^\lambda_k(t),\\
&\dot{\rho}^\lambda_k(t)=\frac{\dot{\omega}_k(t)}{2\omega_k(t)}e^{-2\mi\int^t\omega_k(t')dt'}\gamma^\lambda_k(t).\\
\eea
\ee
The initial no particle state (Bunch-Davis vacuum) can be identified by assuming the condition $\gamma^\lambda_k(t_0)= 1$, $\rho^\lambda_k(t_0) =0$. As time evolves due to time dependent background particles will be produced. In the perturbative limit, at later time we can assume $\rho^\lambda_k(t)<<1$, and $\gamma^\lambda_k(t) \approx 1$. With this approximation, we get
\be
\bea
&\rho^\lambda_k(t)\sim\frac{1}{2}\int^t~dt'\frac{\dot{\omega}_k(t')}{\omega_k(t')}e^{-2\mi\int^{t'}\omega_k(t'')dt''}\\
&=\frac{1}{2}\int^t~\frac{\dot{\omega}_k(t')}{\omega_k(t')}\frac{1}{(-2\mi\omega_k(t')) }
d\Big(e^{-2\mi\Omega_k(t')}\Big)\\
&=\frac{1}{k}\frac{1}{2}\frac{\dot{\omega}_k(t')}{\omega_k(t')}\frac{1}{(-2\mi ) }
\Big(e^{-2\mi\Omega_k(t')}\Big)\Big|^{t}_{t_0}\\
&~~~~~-\frac{1}{k}\int^t~\frac{d}{dt'}\Big[\frac{1}{2}\frac{\dot{\omega}_k(t')}{\omega_k(t')}\frac{1}{(-2\mi)}\Big]
e^{-2\mi\Omega_k(t')}dt'\\
&\simeq\frac{1}{k}\frac{1}{4}\frac{\dot{\mathcal{J}}(t')+\dot{\epsilon}(t')k^2}{\omega^2_k(t')}\frac{1}{(-2\mi ) }
\Big(e^{-2\mi\Omega_k(t')}\Big)\Big|^{t}_{t_0}\\
&~~~~~-\frac{1}{k}\int^t~\frac{d}{dt'}\Big[\frac{1}{4}\frac{\dot{\mathcal{J}}(t')+\dot{\epsilon}(t')k^2}{\omega^2_k(t')}\frac{1}{(-2\mi)}\Big]
e^{-2\mi\Omega_k(t')}dt',\\
\eea
\ee
where we have defined $\Omega_k(t')\equiv \int^{t'}\omega_k(t'')dt''$.
For such integrals, the leading contribution usually comes from the stationary phase approximation. However, we numerically checked that the phase factor $\Omega_k(t')$ does not have any stationary point within the integration limit of one bubble oscillation. In such a case, 
integration by parts leading to a series in $1/k$ in the high frequency limit (the Riemann-Lebesgue lemma, look at page.439 of \cite{Ablowitz}). 
Therefore, the number density in the high frequency limit up to sub-leading order turns out as,
\be
|\beta^\lambda_k(t)|^2\sim|\rho^\lambda_k(t)|^2 \sim  \frac{\dot{\epsilon}^2(t)}{\epsilon^2(t)k^2}+\frac{2\dot{\epsilon}(t)\dot{\mathcal{J}}(t)}{\epsilon^2(t)k^4}e^{\mi\chi(t,k)}+ \frac{\dot{\mathcal{J}}^2(t)}{\epsilon^2k^6} 
 \ee
where, the phase factor, $\chi(t,k)$, arises due to quantum interference \cite{Kolb:2023dzp}.
This immediately leads to the following form of the spectral energy density, 
\be
\frac{\pr\mathcal{E}}{\pr ln k} \sim \frac{\dot{\epsilon}^2(t) k^2}{\epsilon^2(t)}+\frac{2\dot{\epsilon}(t)\dot{\mathcal{J}}(t)}{\epsilon^2(t)}e^{\mi\chi(t,k)}+ \frac{\dot{\mathcal{J}}^2(t)}{\epsilon^2(t)k^2} 
\ee
As stated earlier we indeed see the diverging spectral energy density $\propto k^2$ contribution originated from the time dependent dielectric constant, $\epsilon(t)$ which is due to non-conformal time coordinate. However, if one chooses the conformal time, for example, the case when $\epsilon =1$, leading order spectral energy density falls as $1/k^2$ in large $k$. This is the case we have discussed previously. 

This particular decaying behaviour of the spectral energy density with the frequency, $k$, in the high frequency limit infers that the perturbative approach is not sufficient to produce the required amplitude of the sonoluminescence flux that demands an increasing trend with the frequency up to the ultraviolet range. From our crude estimate, in the experimental range, $k\sim10^7m^{-1}$, spectral energy density turns out as $\frac{\pr\mathcal{E}}{\pr ln k}\sim 10^{-30} \mbox{Watt}/\mbox{nm}$. This estimated amount of flux is much less compared to the nonperturbative production, which we have obtained in the high frequency near the experimentally observed frequency range.

\section{Conclusion} 
Particle production in time dependent background is an intriguing and well-known phenomenon in quantum field theory (QFT) \cite{Schwinger:1951nm, Zeldovich:1971mw, Ford:1986sy}. It has been successfully applied in the early universe cosmological scenario. It would indeed be interesting if such phenomena could be observed in a real laboratory system. Sonoluminescence is one such phenomena where quantum photon production is believed to be one of the possible mechanisms.
A large number of attempts have been made towards this direction in the past with their own pros and cons. In this submission, we have made an attempt to revisit this to understand the photon production mechanism better. We have to say that in the QFT framework if the phenomena is of quantum mechanical origin the underlying mechanism of production should be non-perturbative in nature.

We have modelled the oscillating bubble as an analog geometric system and proposed a nonminimal coupling prescription of the EM field with the analog geometry through the Ricci scalar. Due to oscillating bubble dynamics, the Ricci scalar plays the role of periodic time dependent coupling, and that also breaks the underlying conformal invariance of the EM field. Due to the periodic nature of the Ricci scalar source, and the appropriate values of the coupling parameters, the EM field is observed to show parametric amplification and that can seen as the production of photons flux from the quantum vacuum. Throughout our analysis, we stress the fact that it is this parametric resonance which could be the potential mechanism to explain the observed sonoluminescence phenomena.  

We have computed the photon flux in the parametric resonance regime in the experimental unit. Whereas the experimentally observed frequency is around $k\sim 10^7 \mbox{m}^{-1}$, due to our numerical limitation and high stiffness of the Ricci scalar function, we could obtain the flux in the low frequency region up to $\sim 10^5 {\rm m}^{-1}$. Interestingly magnitude of the produced flux in terms of frequency $(k)$ turned out to be in the required order which may reach the experimental value with good numerical technique. For completeness, we further computed the analytic photon spectrum in the perturbative framework and showed that the produced amplitude of the flux is very low as expected. 

Our model thus suggests that quantum production might be the actual mechanism responsible for photon production in contrast to the existing model on thermal production. Hence, the promising next step would be to analyse the quantumness of the produced spectrum by defining quantum observables, such as Poincare's sphere \cite{Maity:2021zng, Giovanni:2011}, which utilizes time dependent squeezing parameter (one may look at \cite{Maity:2021zng}) for such dynamical systems. Squeezing parameters are connected to the Bogoliuobov coefficients, however, the analysis could be performed in a somewhat, independent manner (see \cite{Martin:2015qta, Chakraborty:2023lpr}), which can also provide for a consistency check of our model. 

In the end, we should mention that the phenomena of sonoluminescence is sensitive to the nature of the gas inside the bubble. In this regard, we want to point out that the theory proposed here is very much dependent on the bubble dynamics including all the parameters (see Sec.\ref{bubble_dynamics}) of the medium inside the bubble. Hence, it will be our next task to take into account different experimental issues in the future \cite{Gaitan, Hiller:1992qz, Gompf}.

\noindent
\textbf{Acknowledgments:} We would like to thank our
Gravity and High Energy Physics groups at IIT Guwahati for fruitful discussions in several contexts. We want to thank Chandramouli Chowdhury for the initial collaboration on this project. Also, we want to thank Swarup Kanti Sarkar for the useful discussion on the Crank-Nicholson method and for providing the relevant references.
\onecolumngrid
\appendix
\section{Derivation of the Rayleigh-Plesset (RP) equation}\label{Rp_eqn}
The Rayleigh-Plesset governing the dynamics of the gas-filled bubble in a medium under acoustic disturbance can be derived in two ways: using the energy balancing between the medium inside and outside of the bubble, and a more advanced way utilising the Navier-Stokes equation. In the following discussion, we will particularly, present the derivation using the approach of energy balancing. For incompressible fluid flow the velocity of the medium can be considered in the form of inverse square law \cite{Rayleigh, Plesset}, given as,     
\be
\frac{u(t,r)}{U(t,r)}=\frac{R^2(t)}{r^2},
\ee
where, $U(t,r)$ denotes the velocity at the surface of the bubble and $u(t,r)$ represents the velocity at $r>R$. With the consideration, $U\sim \dot{R}$, we obtain
\be
u(t,r)=\frac{R^2(t)}{r^2}\dot{R}(t).
\ee
 By equating the kinetic energy of the fluid with the work done by the net pressure on the bubble surface, we arrive at
\be
\bea
&\frac{1}{2}\rho\int^R_{r=\infty} 4\pi r^2 u^2 dr =\int \left( P(R) - P_0 + P_a(t) + \frac{R}{c_s} \d{}{t}[P_g(R) + P_a(t)]\right) 4\pi R^2 d R,\\
\implies&2\pi \rho R^3\dot{R}^2=\int \left( P(R) - P_0 + P_a(t) + \frac{R}{c_s} \d{}{t}[P_g(R) + P_a(t)]\right) 4\pi R^2 d R,\\
\eea
\ee
where all the physical quantities have been described in Sec.\ref{bubble_dynamics}. Now integrating both sides of the above equation with respect to $R(t)$, we get,
\be 
\bea
&R\ddot{R}+\frac{3}{2} \dot{R}^2= \frac{1}{\rho} \left(P(R) - P_0 + P_a(t) + \frac{R}{c_s} \d{}{t}[P_g(R) + P_a(t)]\right).
\eea
\ee
\section{Diagonalization Of The Metric}\label{diagonal}
The off-diagonal fluctuation metric discussed in the text is given by 
\be
ds^2=-f(t,r)dt^2+2g(t,r)drdt+p(t)\left(dr^2+r^2d\Omega^2\right).
\ee
Now we rewrite the above metric to make it suitable for diagonalization,
\be
ds^2=-f(t,r)dt^2-\frac{g^2(t,r)}{p(t)}dt^2+\left(\sqrt{p(t)} \,d r + \frac{g(t,r)}{\sqrt{p(t)}} \,dt\right)^2+p(t)r^2d\Omega^2.
\ee
Considering the following transformation in the radial direction (look at Chapter-7 of \cite{Padmanabhan:2010zzb} for discussions on coordinate transformation to diagonalize a spherically symmetric metric),
\be
\frac{\,d \bar{r}}{X(t,r)} = \sqrt{p(t)} \,d r + \frac{g(t,r)}{\sqrt{p(t)}} \,dt.
\ee
Where $X(t,r)$ some arbitrary function, will be found out in a moment. This transformation leads us to deduce the following relations:
\be
\bea
&\pd{\bar{r}}{r} =X(t,r)\sqrt{p(t)}, \\
&\pd{\bar{r}}{t} = X(t,r)\frac{g(t,r)}{\sqrt{p(t)}}.
\eea
\ee
Recall the metric coefficients $p(t)$ and $g(t,r)$ given in the main text as,
\be
\bea
&p(t) = 1 + \frac{\xi^3}{R^3},\\
&g(r, t) = - \frac{\dot{R} \xi^3}{R^4} r.
\eea
\ee
Denoting, $\sigma(t) =\xi^3/(R^3)$, we have the following integral equations respectively, 
\be
\bea
&\bar{r} = \frac{r}{3} \int \,d \sigma \frac{X(t,r)}{\sqrt{1 + \sigma}} +\gamma(r),\\
&\bar{r} = \sqrt{1 + \sigma(t)} \int \,dr X(t,r) +\tilde{\gamma}(t).
\eea
\ee
To simplify our task we proceed by considering  $X(t,r)\equiv \sim X(\sigma(t))$, i.e. dependence of $X$ only on time. This gives us the following equation for $\bar{r}$, 
\be\label{tildersigma}
\bar{r} = \frac{r}{3} \int \,d \sigma \frac{\zeta(\sigma)}{\sqrt{1 + \sigma}}=r\sqrt{1 + \sigma}X(\sigma),
\ee
where, we have set $\gamma(r)=\tilde{\gamma}(t)=0$. Differentiating the later two equations w.r.t $\sigma$ gives us the following differential equation for $X$, 
\be
-\frac{1}{2}X = 3 (1 + \sigma) \d{X}{\sigma}.
\ee
Integrating both sides of the above equation, we arrive at, 
\be
X(\sigma) = \frac{1}{(1 + \sigma)^{1/6}}.
\ee
Now, we know the functional form of the arbitrary function we assumed. Substituting the function in \eqref{tildersigma} we finally obtain,
\be
\bar{r} = r ( 1 + \sigma)^{1/3}=r p^{1/3}.
\ee
This rescaling of the radial coordinate will help one to diagonalize the metric. 
\section{Discretization of second order ODE using Crank-Nicholson Method}\label{CN}
Having stiffness in the differential equation, arising from the sudden collapse of the water bubble, we resort to using those numerical discretization methods, which provide for better stability. The crank-Nicholson method is a popular implicit method \cite{Bao, Adhikari}, mostly used in the case of partial differential equations, such as the heat equation, to discretize the first order differentiation in the time part. Generically, one can consider the following model differential equation   
\be
P(x)\frac{d^2y}{dx^2}+Q(x)\frac{dy}{dx}+R(x)y=0,
\ee
with the initial conditions given as, $y(x_0)$ and $y_1(x_0)$. Now, we decompose the above second order ODE into two first order ODE,
\be\label{redefine}
\frac{dy}{dx}=y_1,~~\frac{dy_1}{dx}=-\frac{Q(x)y_1+R(x)y}{P(x)}.
\ee
The crank-Nicholson  method is attributed to the following discretization scheme,
\be\label{CN_form}
\bea
y^{n+1}&=y^n+\frac{h}{2}\left(\frac{dy}{dx}\Big|_{x_n}+\frac{dy}{dx}\Big|_{x_n+h}\right),\\
&=y^n+\frac{h}{2}\left(y_1|_{x_n}+y_1|_{x_n+h}\right),\\
\eea
\ee
which we can also apply to the second equation of \eqref{redefine} to obtain,
\be
\bea
y^{n+1}_1&=y^n_1+\frac{h}{2}\left(\frac{dy_1}{dx}\Big|_{x_n}+\frac{dy_1}{dx}\Big|_{x_n+h}\right).\\
\eea
\ee
Where $n\in \mathcal{Z}$, set of integers, and $y^n\equiv y(x_n)$, $y_1^n\equiv y_1(x_n)$. Using the explicit expressions for $dy_1/dx$ from \eqref{redefine}, we can rewrite the above equation as,
\be
y^{n+1}_1=y^n_1-\frac{h}{2}\left(\frac{dy_1}{dx}\Big|_{x_n}+\frac{Q(x_n+h)y^{n+1}_1+R(x_n+h)y^{n+1}}{P(x_n+h)}\right).
\ee
Now substituting the discretization of $y^{n+1}$ from \eqref{CN_form} we obtain,
\be
\bea
&y^{n+1}_1=y^n_1+\frac{h}{2}\left[\frac{dy_1}{dx}\Big|_{x_n}-\frac{Q(x_n+h)y^{n+1}_1}{P(x_n+h)}-\frac{R(x_n+h)}{P(x_n+h)}\left\{y^n+\frac{h}{2}\left(y_1\Big|_{x_n}+y_1\Big|_{x_n+h}\right)\right\}\right],\\
\implies& y^{n+1}_1=\frac{y^n_1\left\{1-\left(\frac{h}{2}\right)^2\frac{R(x_n+h)}{P(x_n+h)}\right\}+\frac{h}{2}\left\{\frac{dy_1}{dx}\Big|_{x_n}-\frac{R(x_n+h)}{P(x_n+h)}y_n\right\}}{\left[1+\frac{h}{2}\frac{Q(x_n+h)}{P(x_n+h)}+\left(\frac{h}{2}\right)^2\frac{R(x_n+h)}{P(x_n+h)}\right]} .
\eea
\ee
In the second step of the above equation, we have rearranged the terms, such that all the known quantities stay on the right-hand side. The steps up to the evaluation of the solution proceed as follows: starting from the initial conditions $y(x_0)$ and $y_1(x_0)$, evaluate the $y_1(x_0+h)$ from the above equation, and substitute in \eqref{CN_form}, which will provide for the solution $y(x_0+h)$. Now repeat the iteration to generate the solution for the consecutive points.  
\section{Fitted polynomial for the number density}\label{fitfunc}
Having a promising trend (fig.\ref{betaksqrwk}) in the number density of photons, evaluated through our formalism, with the experimental measurement, compelled us to fit our results by utilizing the following polynomial, 
\be
\bea
y=&93.6 x - 105.5 x^2 + 51.3 x^3 - 12.6 x^4 + 1.6 x^5 - 
 0.08 x^6 - 0.004 x^7 + 0.0008 x^8\\
&- 5\times10^{-6} x^9 + 1.4 \times10^{-6}x^{10} -1.6 \times10^{-8} x^{11},
\eea
\ee
where $x\equiv\ln k$ and $y\equiv\ln |\beta_k|^2$.

\end{document}